\documentclass[universe,review,accept,moreauthors,pdftex]{Definitions/mdpi} 

\firstpage{1} 
\pubvolume{8}
\issuenum{1}
\articlenumber{169}
\pubyear{2022}
\copyrightyear{2022}
\externaleditor{Academic Editor: {Bertrand Echenard and Robert H. Bernstein } 
} 
\datereceived{20 December 2021} 
\dateaccepted{4 March 2022} 
\datepublished{8 March 2022} 
\hreflink{https://doi.org/ \\
10.3390/universe8030169} 

\Title{Studying $\Delta L=2$ Lepton Flavor Violation with Muons}

\TitleCitation{Studying $\Delta L=2$ Lepton Flavor Violation with Muons}

\Author{Alexey A. Petrov 
 $^{1,2,3,}$*
\orcidB{}, Renae 
 Conlin $^{1}$\orcidA{} and Cody Grant $^{1}$}

\AuthorNames{Alexey A. Petrov, Renae Conlin and Cody Grant}

\AuthorCitation{Petrov, A.A.; Conlin, R.; Grant, C.}

\address{%
$^{1}$ \quad Department of Physics and Astronomy, 
 Wayne State University, Detroit, MI 48201, USA; renae.cederquist@wayne.edu (R.C.); codygrant@wayne.edu (C.G.)\\
$^{2}$ \quad Theoretical Physics Department, Fermilab,
P.O. Box 500, Batavia, IL 60510, USA \\
$^{3}$ \quad Excellence Cluster ORIGINS, 
Technische Universit\"at Mu\"nchen, D-85748 Garching, 
 Germany}

\corres{Correspondence: apetrov@wayne.edu}

\abstract{Flavor violating processes in the lepton sector have highly suppressed branching ratios in the 
standard model. Thus, observation of lepton flavor violation (LFV) constitutes a clear indication of 
physics beyond the standard model (BSM). We review new physics searches in the processes that 
violate the conservation of lepton (muon) flavor by two units with muonia and muonium--antimuonium 
oscillations. 
}


\newcommand{\ket}[1]{\left|#1\right\rangle}
\newcommand{\bra}[1]{\left\langle#1\right|}

\newcommand{\beq}{\begin{equation}}
\newcommand{\eeq}{\end{equation}}
\newcommand{\beqa}{\begin{eqnarray}}
\newcommand{\eeqa}{\end{eqnarray}}

\newcommand{\mmbar}{M_\mu-\overline{M}_\mu}
\newcommand{\mmu}{{M_\mu}}
\newcommand{\mmup}{{M_\mu^P}}
\newcommand{\mmuv}{{M_\mu^V}}
\newcommand{\ammu}{{\overline{M}_\mu}}
\newcommand{\ammup}{{\bar{M}_\mu^P}}
\newcommand{\ammuv}{{\bar{M}_\mu^V}}

\usepackage{slashed,simplewick}

\begin{document}

\section{Introduction}\label{Intro}

Flavor-changing neutral current (FCNC) interactions serve as a powerful probe of physics beyond the standard 
model (BSM). Since no local operators generate FCNCs in the standard model (SM) at tree level, new physics (NP) degrees of 
freedom can effectively compete with the SM particles running in the loop graphs, making their discovery possible.
This is, of course, only true provided the BSM models include flavor-violating interactions. If the new physics particles are 
heavier than the muon mass, their effect on muon transitions can be parameterized in terms of local operators of increasing
dimension. In fact, any new physics scenario which involves lepton flavor violating interactions can be matched to an effective 
Lagrangian of the Standard Model Effective Field Theory (SM EFT), 
\beq\label{lagrang}
{\cal L}_{\rm SM EFT}=-\frac{1}{\Lambda^{2}}\underset{i}{\sum}c_{i}(\mu)Q_{i},
\eeq
where the Wilson coefficients (WC) $c_{i}$ of ${\cal L}_{\rm SM EFT}$ in Equation (\ref{lagrang}) are determined by the 
UV physics that becomes active at some scale $\Lambda$ \cite{Petrov:2021idw,Grzadkowski:2010es,Petrov:2016azi}. 

The effective operators $Q_{i}$'s defined in Equation (\ref{lagrang}) reflect degrees of freedom relevant at the scale at which a 
given process takes place.
For heavy new physics scenarios, those operators should be written in terms of muon, electron, neutrino, or 
photon (gluon) fields. Then to perform low energy computations, it is often convenient to match the SM EFT Lagrangian to the 
low energy Lagrangian ${\cal L}_{\rm eff}$ only containing these degrees of freedom. In addition, it is convenient to classify the 
operators of ${\cal L}_{\rm eff}$ by their lepton quantum numbers $L_\ell$ (with $\ell = e,\mu,\tau$),
\beq\label{LeptNumbLagr}
{\cal L}_{\rm eff} = {\cal L}_{\rm eff}^{\Delta L_\mu=0} + {\cal L}_{\rm eff}^{\Delta L_\mu=1} + {\cal L}_{\rm eff}^{\Delta L_\mu=2}
\eeq

We shall employ the operators that only contain the fermion fields, so the operators in Equation (\ref{LeptNumbLagr}) start at dimension six.

The first term in Equation (\ref{LeptNumbLagr}) contains both the standard model and the new physics contributions. The SM piece is 
given by the usual Fermi Lagrangian,
\beq\label{SMLagr}
{\cal L}_{\rm eff}^{\Delta L_\mu=0} = -\frac{4 G_F}{\sqrt{2}} 
\left(\overline\mu_L \gamma_\alpha e_L \right) \left(\overline{\nu_e}_L \gamma^\alpha {\nu_\mu}_L \right).
\eeq

Since $\Lambda \gg m_W$, which is required for the consistency of SM EFT, we can neglect the NP piece of the 
$\Delta L_\mu = 0$ Lagrangian for studies of FCNC transitions.
This implies that ${\cal L}_{\rm eff}^{\Delta L_\mu=0}$ is suppressed by powers of $G_F \sim M_W^{-2}$, the Fermi constant. 
We should emphasize that only the operators that are local at the scale of the muonium mass are retained in 
Equation (\ref{LeptNumbLagr}).

The second term in Equation (\ref{LeptNumbLagr}) contains $\Delta L_{\mu} = 1$ operators. 
The most general low energy Lagrangian representing muon flavor change by one unit, $\mathcal{L}_{\rm eff}^{\Delta L_{\mu}=1}$, 
is conventionnaly written as \cite{Celis:2014asa,Hazard:2017udp,Hazard:2016fnc}
\begin{eqnarray}\label{L_eff1}
{\cal L}_{\rm eff}^{\Delta L_{\mu}=1} = &-&\frac{1}{\Lambda^2} \sum_f \Big[
\left( C_{VR}^{f} \ \overline\mu_R \gamma^\alpha e_R + 
C_{VL}^{f} \ \overline\mu_L \gamma^\alpha e_L \right) \ \overline f \gamma_\alpha f 
\nonumber \\
&+& \
\left( C_{AR}^{f} \ \overline\mu_R \gamma^\alpha e_R + 
C_{AL}^{q} \ \overline\mu_L \gamma^\alpha e_L \right) \ \overline f \gamma_\alpha \gamma_5 f 
\nonumber \\
&+& \
m_e m_f G_F \left( C_{SR}^{f} \ \overline\mu_R e_L + 
C_{SL}^{f} \ \overline\mu_L e_R \right) \ \overline f f 
\\
&+& \
m_e m_f G_F \left( C_{PR}^{f} \ \overline\mu_R e_L + 
C_{PL}^{f} \ \overline\mu_L e_R \right) \ \overline f \gamma_5 f
\nonumber \\
&+& \
m_e m_f G_F \left( C_{TR}^{f} \ \overline\mu_R \sigma^{\alpha\beta} e_L + 
C_{TL}^{f} \ \overline\mu_L \sigma^{\alpha\beta} e_R \right) \ \overline f \sigma_{\alpha\beta} f 
 + h.c.   \Big],
\nonumber
\end{eqnarray}
where $\mu$ and $e$ are the fermion fields with $(\mu,e)_{L,R} = P_{L,R}(\mu,e)$, where 
$P_{R,L} = \frac{1}{2}\left(1\pm\gamma^{5}\right)$ are the projection operators, and $f$ represents other 
fermions that are not integrated out from the low energy effective theory. We introduced indices for the 
Wilson coefficients $C_{IK}^{f}$ to indicate the Lorentz structure of effective operators: vector, axial-vector, scalar, 
pseudo-scalar, and tensor. The Wilson coefficients are in general different for different fermions $f$.
The operators additionally suppressed by factors of $m_i m_j G_F$ usually arise from SM EFT operators of 
dimension higher than six \cite{Celis:2014asa,Hazard:2017udp}.

The last term in the low energy EFT described in Equation (\ref{LeptNumbLagr}), ${\cal L}_{\rm eff}^{\Delta L_\mu=2}$, 
represents a collection of effective operators changing the lepton quantum number by two units. The most general 
effective Lagrangian for such transition that is applicable to the physics we will discuss in this review is given by 
\begin{equation}\label{DL2}
{\cal L}_{\rm eff}^{\Delta L_\mu=2}=-\frac{1}{\Lambda^{2}}\underset{i}{\sum}C^{\Delta L=2}_{i}(\mu)Q_{i}(\mu).
\end{equation}
with the operators written entirely in terms of the muon and electron degrees of freedom,
\begin{eqnarray}\label{Dim6_Op}
Q_1 &=& \left(\overline\mu_L \gamma_\alpha e_L \right) \left(\overline\mu_L \gamma^\alpha e_L \right), \quad
Q_2 = \left(\overline\mu_R \gamma_\alpha e_R \right) \left(\overline\mu_R \gamma^\alpha e_R \right),
\nonumber \\
Q_3 &=& \left(\overline\mu_L \gamma_\alpha e_L \right) \left(\overline\mu_R \gamma^\alpha e_R \right), \quad	
Q_4 = \left(\overline\mu_L e_R \right) \left(\overline\mu_L e_R \right), 
\nonumber \\
Q_5 &=& \left(\overline\mu_R e_L \right) \left(\overline\mu_R e_L \right).
\end{eqnarray}

Other Lorentz structures of the operators could be related to the ones in Equation (\ref{Dim6_Op}) via Fierz relations. 
A $\Delta L=2$ interaction described by this Lagrangian can lead to muon decays such as $\mu^+ \to 3e$. 
It can also change a $\mu^+ e^-$ bound state into a $\mu^- e^+$ bound state, leading to muonium--antimuonium 
oscillations. We will discuss those below.

In addition, the operators built out of muon and electron fields, we can construct other $\Delta L_\mu=2$ local operators 
that contain muon, electron, and neutrino fields \cite{Conlin:2020veq},
\beq\label{Dim6_Op_nu}
Q_6 = \left(\overline\mu_L \gamma_\alpha e_L \right) \left(\overline{\nu_{\mu}}_L \gamma^\alpha {\nu_e}_L \right), \quad
Q_7 = \left(\overline\mu_R \gamma_\alpha e_R \right) \left(\overline{\nu_{\mu}}_L \gamma^\alpha {\nu_e}_L \right),
\eeq
where we only included operators that contain left-handed neutrinos \cite{Petrov:2016azi,Grossman:2003rw}. The operators 
in Equation (\ref{Dim6_Op_nu}) could lead to both muon decays $\mu \to e \nu_\mu \bar \nu_e$ (note the ``reversed" neutrino flavors),
and muonium--antimuonium oscillations. 

The goal of experimental studies of $\Delta L=2$ transitions would be to find the observables that are sensitive to 
various combinations of Wilson coefficients of the effective Lagrangian in Equation (\ref{LeptNumbLagr}). Different processes 
are in general sensitive to different combinations of the Wilson coefficients, which raises hope that a sufficient number of 
measurements would allow placing constraints on individual WCs without additional assumptions such as single operator 
dominance \cite{Hazard:2016fnc,Hazard:2017udp}. In that respect, studies of both unbound muon and muonium decays 
are needed.
 
\section{Muonium: The Simplest Bound State}

Even though the hydrogen atom is a quintessential quantum-mechanical bound state and is usually presented as the 
easiest QED problem, it is not the simplest. Precision studies of hydrogen reveal effects related to the structure of the 
proton: its finite size and composite nature. In that respect, the simplest hydrogen-like system is built entirely out of 
leptons. Such a system that is especially clean for studies of BSM effects in the lepton sector is muonium $\mmu$. The muonium 
is a QED bound state of a positively-charged muon and a negatively-charged electron, $|\mmu\rangle \equiv |\mu^+e^-\rangle$. 

Since muon is unstable, $\mmu$ is unstable. The main decay channel for the muonium is determined by the weak decay of the 
muon, $\mmu \to e^+e^- \bar \nu_\mu \nu_e$, so the average lifetime of a muonium state $\tau_\mmu$ is 
expected to be the same as that of the muon, 
\beq\label{MuoniumWidth}
1/\tau_\mmu = \Gamma\left(\mmu \to e^+e^- \nu_e \bar\nu_\mu \right) \approx
\Gamma\left(\mu \to e^+ \nu_e \bar\nu_\mu \right) = \frac{G_F^2 m_\mu^5}{192 \pi^3}
=1/\tau_\mu , 
\eeq
with $\tau_\mu = (2.1969811 \pm 0.0000022)\times 10^{-6}$ s \cite{Tanabashi:2018oca},
apart from the tiny effect due to time dilation \cite{Czarnecki:1999yj}. Note that Equation (\ref{MuoniumWidth}) represents
the leading-order result. The results including subleading corrections are available \cite{Czarnecki:1999yj}.

Like a hydrogen atom, muonium could be formed in two spin configurations. A spin-one triplet state $\ket{\mmuv}$ is
called {\it ortho-muonium}, 
while a spin-zero singlet state $\ket{\mmup}$ called {\it para-muonium}. In what follows, we 
will drop the superscript and employ the notation $\ket{\mmu}$ if the spin of the muonium state is not important for the 
discussion.

\section{Muonium--Antimuonium Oscillations}

Since $\Delta L=2$ interaction can change the muonium state into the anti-muonium one, the possibility to study 
muonium--anti-muonium oscillations arises. Theoretical analyses of conversion probability for muonium into 
antimuonium have been performed, both in particular new physics
models \cite{Pontecorvo:1957cp,Feinberg:1961zza,ClarkLove:2004,CDKK:2005,Li:2019xvv,Endo:2020mev}, and 
using the framework of effective theory \cite{Conlin:2020veq}, where all possible BSM models are encoded 
in a few Wilson coefficients of effective operators. Observation of muonium converting into anti-muonium provides clean 
probes of new physics in the leptonic sector \cite{Bernstein:2013hba,Willmann:1998gd}. 

\subsection{Phenomenology of Muonium Oscillations}\label{Pheno}

In order to determine experimental observables related to $\mmbar$ oscillations, we recall that 
the treatment of the two-level system that represents muonium and antimuonium is similar to that 
of meson-antimeson oscillations \cite{Petrov:2021idw,Donoghue,Nierste}. There are, however, several
important differences. First, both ortho- and para-muonium can oscillate. Second, the SM oscillation
probability is tiny, as it is related to a function of neutrino masses, so any experimental indication of oscillation
would represent a sign of new physics. 
 
In the presence of the interactions coupling $\mmu$ and $\ammu$, the time development of a muonium 
and anti-muonium states would be coupled, so it would be appropriate to consider their combined evolution, 
\beq
|\psi(t)\rangle = 
\left( {\begin{array}{c}
 a(t) \\
 b(t) \\
 \end{array} } \right) =
 a(t) |\mmu\rangle + b(t) |\ammu\rangle.
\eeq

The time evolution of $|\psi(t)\rangle$ evolution is governed by a Schr\"odinger-like equation, 
\begin{equation}\label{two state time evolution}
i\frac{d}{dt}
\begin{pmatrix}
\ket{M_\mu(t)} \\ \ket{\overline{M}_\mu(t)}
\end{pmatrix}
=
\left(m-i\frac{\Gamma}{2}\right)
\begin{pmatrix}
\ket{M_\mu(t)} \\ \ket{\overline{M}_\mu(t)}
\end{pmatrix}.
\end{equation}
where $\left(m-i\frac{\Gamma}{2}\right)_{ik}$ is a $2\times 2$ Hamiltonian (mass matrix) with non-zero off-diagonal
terms originating from the $\Delta L=2$ interactions. CPT-invariance dictates that the masses and widths of 
the muonium and anti-muonium are the same, so $m_{11}= m_{22}$, $\Gamma_{11}=\Gamma_{22}$.
In what follows, we assume CP-invariance of the $\Delta L_\mu = 2$ interaction\endnote{A more general formalism 
without this assumption follows the same steps as that for the $B\bar B$ or $K\bar K$ mixing \cite{Petrov:2021idw,Donoghue}.}.
Then,
\begin{eqnarray}\label{off_diagonal_elements}
m_{12}=m^{*}_{21}, \qquad \Gamma_{12}=\Gamma^{*}_{21}.
\end{eqnarray}

The off-diagonal matrix elements in Equation (\ref{off_diagonal_elements}) can be related to the matrix elements of 
the effective operators introduced in Section \ref{Intro}, as discussed in \cite{Petrov:2021idw,Donoghue},
\begin{equation}\label{OffDiagonal}
\left(m-\frac{i}{2} \Gamma\right)_{12}=\frac{1}{2 M_{M}}\left\langle\ammu\left|{\cal H}_{\rm eff} \right| \mmu\right\rangle
+\frac{1}{2 M_{M}} \sum_{n} \frac{\left\langle\ammu
\left|{\cal H}_{\rm eff} \right| n\right\rangle\left\langle n\left|{\cal H}_{\rm eff} \right| \mmu\right\rangle}{M_{M}-E_{n}+i \epsilon}.
\end{equation}

To find the propagating states, the mass matrix needs to be diagonalized. The basis in which the 
mass matrix is diagonal is represented by the mass eigenstates $ |\mmu_{1,2} \rangle$, which are 
related to the flavor eigenstates $\mmu$ and $\ammu$ as
\beq
 |\mmu_{1,2} \rangle = \frac{1}{\sqrt{2}} \left[ |\mmu \rangle \mp |\ammu \rangle
 \right] ,
\eeq
where we employed a convention where $CP |\mmu_\pm \rangle = \mp |\mmu_\pm \rangle$.
The mass and the width differences of the mass eigenstates are 
\begin{eqnarray}\label{def of mass and width difference}
\Delta m \equiv M_{1}-M_{2}, \qquad \Delta \Gamma \equiv \Gamma_{2}-\Gamma_{1}.
\end{eqnarray} 

Here, $M_i$ ($\Gamma_i$) are the masses (widths) of the physical mass eigenstates $ |\mmu_{1,2} \rangle$. 

It is interesting to see how the Equation (\ref{OffDiagonal}) defines the mass and the lifetime differences.
Since the first term in Equation (\ref{OffDiagonal}) is defined by a local operator, its matrix element 
does not develop an absorptive part, so it contributes to $m_{12}$, i.e., the mass difference. 
The second term contains bi-local contributions connected by physical intermediate states. This term 
has both real and imaginary parts and thus contributes to both $m_{12}$ and $\Gamma_{12}$.

It is often convenient to introduce dimensionless quantities,
\beq\label{XandY}
x = \frac{\Delta m}{\Gamma}, \qquad y = \frac{\Delta \Gamma}{2\Gamma},
\eeq
where the average lifetime $\Gamma=(\Gamma_1+\Gamma_2)/2$, and $M_M = \left(M_1+M_2\right)/2$ 
is the muonium mass. Noting that $\Gamma$ is defined by the standard model decay rate of the muon, 
and $x$ and $y$ are driven by the lepton-flavor violating interactions, we should expect that both $x,y \ll 1$.

The time evolution of flavor eigenstates follows from Equation (\ref{two state time evolution}) \cite{Donoghue,Nierste,Conlin:2020veq},
\beqa
\ket{M(t)} &=& g_+(t) \ket{\mmu} + g_-(t) \ket{\ammu},
\nonumber \\
\ket{\overline{M}(t)} &=& g_-(t) \ket{\mmu} + g_+(t) \ket{\ammu},
\eeqa
where the coefficients $g_\pm(t)$ are defined as
\beq\label{TimeDep}
g_\pm(t) = \frac{1}{2} e^{-\Gamma_1 t/2} e^{-iM_1t} \left[1 \pm e^{\Delta\Gamma t/2} e^{i \Delta m t} \right].
\eeq

As $x,y \ll 1$ we can expand Equation (\ref{TimeDep}) in power series in $x$ and $y$ to obtain
\beqa
g_+(t) &=& e^{-\Gamma_1 t/2} e^{-iM_1t} \left[1 + \frac{1}{8} \left(y-ix\right)^2 \left(\Gamma t\right)^2 \right],
\nonumber \\
g_-(t) &=& \frac{1}{2} e^{-\Gamma_1 t/2} e^{-iM_1t} \left(y-ix\right) \left(\Gamma t\right).
\eeqa

The most natural way to detect $\mmbar$ oscillations experimentally is by producing $\mmu$ state and looking for the decay 
products of the CP-conjugated state $\ammu$. Denoting an amplitude for the $\mmu$ decay into a final state 
$f$ as $A_f = \langle f|{\cal H} |\mmu\rangle$ and an amplitude for its decay into a CP-conjugated final state 
$\overline{f}$ as $A_{\bar f} = \langle \overline{f}|{\cal H} |\mmu\rangle$,
we can write the time-dependent decay rate of $\mmu$ into the $\overline{f}$,
\beq
\Gamma(\mmu \to \overline{f})(t) = \frac{1}{2} N_f \left|A_f\right|^2 e^{-\Gamma t} \left(\Gamma t\right)^2 R_M(x,y),
\eeq
where $N_f$ is a phase-space factor and we defined the oscillation rate $R_M(x,y)$ as
\beq
R_M(x,y) = \frac{1}{2} \left(x^2+y^2\right).
\eeq

Integrating over time and normalizing to $\Gamma(\mmu \to f)$ we get the probability of $\mmu$ decaying as $\ammu$ at some time $t > 0$,
\beq\label{Prob_osc}
P(\mmu \rightarrow \ammu) = \frac{\Gamma(\mmu \to \overline{f})}{\Gamma(\mmu \to f)} = R_M(x,y). 
\eeq

The equation Equation (\ref{Prob_osc}) \cite{Conlin:2020veq} generalizes oscillation probability found in the 
papers \cite{Feinberg:1961zza,CDKK:2005} by allowing for a non-zero lifetime difference in $\mmbar$ oscillations.
We will review how $x$ and $y$ are related to the fundamental parameters of the Lagrangian below \cite{Conlin:2020veq}. 

\subsection{The Mass Difference $x$}\label{MassDiff}

The physical mixing parameters $x$ and $y$ can be obtained from Equation (\ref{XandY}). 
The mass difference $x$ comes from the dispersive part of the correlator.
\beq\label{MEforX}
x = \frac{1}{2 M_M \Gamma} \mbox{Re} \left[ 
2 \langle\ammu\left|{\cal H}_{\rm eff} \right| \mmu\rangle 
+ \langle\ammu \left| i\int d^4x \ \mbox{T} \left[
 {\cal H}_{\rm eff} (x) {\cal H}_{\rm eff} (0) \right] \right | \mmu\rangle 
\right].
\eeq

Neglecting the SM contribution to the local $\Delta L_\mu = 2$ Hamiltonian, 
$\langle \ammu | {\cal H}_{\rm eff} |\mmu\rangle = \langle \ammu | {\cal H}_{\rm eff}^{\Delta L_\mu=2} |\mmu\rangle$,
so the dominant contribution is only suppressed by $\Lambda^2$. We can neglect the contribution of the second term in 
Equation (\ref{MEforX}), as it is suppressed by either $\Lambda^4$ or by $M_W^2 \Lambda^2$. In the former case this 
suppression comes from the double insertion of the ${\cal L}_{\rm eff}^{\Delta L_{\mu}=1}$ term, each of which is 
suppressed by $\Lambda^2$, as follows from Equation (\ref{L_eff1}). In the later, the suppression comes from the 
insertion of the SM ${\cal L}_{\rm eff}^{\Delta L_{\mu}=0}$ and ${\cal L}_{\rm eff}^{\Delta L_{\mu}=2}$ terms with the
operators given in Equation (\ref{Dim6_Op_nu}).

Computation of the mass and the lifetime differences involves evaluating the matrix elements between the 
muonium states for both the spin-0 singlet and the spin-1 triplet configurations. Since $\mmu$ is a QED bound state,
such matrix elements can be written in terms of the value of the muonium wave function at the origin. 

In the non-relativistic approximation, which is applicable for the muonium, it is given by a Coulombic bound state 
the wave function of the ground state, 
\beq
\varphi (r)=\frac{1}{\sqrt{\pi a_{\tiny \mmu}^{3}}}e^{-\frac{r}{a_{\tiny \mmu}}}, 
\eeq
where $a_{\tiny \mmu}=(\alpha m_{\rm red})^{-1}$ is the muonium Bohr radius, $\alpha$ is the fine structure constant, 
and $m_{red}=m_{e}m_{\mu}/(m_{e}+m_{\mu})$ is the reduced mass. The absolute value $|\varphi(0)|$ at the origin
can be written as
\begin{equation}\label{WFat0}
|\varphi(0)|^2=\frac{(m_{red}\alpha)^{3}}{\pi} =\frac{1}{\pi} (m_{red}\alpha)^3,
\end{equation} 
where we substituted the value of the Bohr radius. The applicability of a non-relativistic approximation to muonium can be 
established from a simple scaling argument. The typical momentum in the muonium state is
\begin{equation}\label{NonRelMu}
p \simeq \frac{\hbar}{a_{\tiny \mmu}} = \frac{m_{\rm red} e^2}{\hbar} = m_{\rm red} \left(\frac{e^2}{\hbar c} \right) c \equiv m_{\rm red} v 
\end{equation}
where, for a moment, we reinstated $c$ and $\hbar$. 
We can see from Equation (\ref{NonRelMu}) that $v \sim \alpha c$, justifying the non-relativistic approximation. 

However, it might be easier to apply a factorization approach familiar from the description of meson flavor oscillations.
In this approach the matrix elements in Equation (\ref{MEforX}) are obtained by inserting a vacuum state to turn 
matrix elements of four-fermion operators $\bra{\ammu} ... \ket{\mmu}$ into products 
$\bra{\ammu} ... \ket{0} \bra{0} ... \ket{\mmu}$ of matrix elements of current operators. Such matrix elements 
can be further parameterized as 
\begin{eqnarray}\label{DecayConst}
\bra{0} \overline{\mu} \gamma^{\alpha} \gamma^{5} e \ket{\mmup} &=& i f_P p^{\alpha}, \quad
\bra{0} \overline{\mu} \gamma^\alpha e \ket{\mmuv} = f_V M_{M} \epsilon^\alpha (p), 
\nonumber \\
\bra{0} \overline{\mu} \sigma^{\alpha\beta} e \ket{\mmuv} &=& i f_T \left(\epsilon^\alpha p^\beta - \epsilon^\beta p^\alpha\right),
\end{eqnarray}
where $f_{M}$ is the muonium decay constant \cite{Conlin:2020veq,Hazard:2016fnc}, $p^\alpha$ is muonium's four-momentum, 
and $\epsilon^\alpha (p)$ is the ortho-muonium's polarization vector.
Note that $f_P=f_V=f_T=f_M$ in the non-relativistic limit. The decay constant $f_{M}$ can be expressed in terms of the 
bound-state wave function using the QED version of Van Royen-Weisskopf formula,
\begin{eqnarray}\label{Definition_fM}
f_M^2 = 4\frac{\left|\varphi(0)\right|^2}{M_{M}}. 
\end{eqnarray}

This factorization gives the exact result for the QED matrix elements of the six-fermion operators in the non-relativistic limit,
as can be explicitly verified \cite{Conlin:2020veq}. Note that muonium mass and lifetime differences, and decay probabilities are
thus suppressed by $|\varphi(0)|^2 \sim m_e^3$.

{\bf Para-muonium.} 
 The matrix elements of the spin-singlet states can be obtained from Equation (\ref{Dim6_Op})
using the definitions of Equation (\ref{DecayConst}),
\begin{eqnarray}\label{ME0}
 \bra{\ammup} Q_{1}\ket{\mmup} &=& \ \ f_M^2 M_M^2, \ \quad
 \bra{\ammup} Q_{2}\ket{\mmup} = \ \ f_M^2 M_M^2, \nonumber \\ 
 \bra{\ammup} Q_{3}\ket{\mmup} &=& -\frac{3}{2} f_M^2 M_M^2, \quad
 \bra{\ammup} Q_{4}\ket{\mmup} = - \frac{1}{4} f_M^2 M_M^2, \nonumber \\
 \bra{\ammup} Q_{5}\ket{\mmup} &=& - \frac{1}{4} f_M^2 M_M^2.
\end{eqnarray}

Combining the contributions from the different operators and using the definitions from Equations (\ref{WFat0}) and (\ref{Definition_fM}), 
we obtain an expression for $x_P$ for the para-muonium state,
\begin{equation}\label{Delta_m_para}
x_P=\frac{4 (m_{red}\alpha)^{3}}{\pi\Lambda^2 \Gamma} 
\left[
C_1^{\Delta L= 2} + C_2^{\Delta L= 2} - \frac{3}{2} C_3^{\Delta L= 2} - 
\frac{1}{4}\left(C_4^{\Delta L= 2} + C_5^{\Delta L= 2}\right)
\right] .
\end{equation}

{\bf Ortho-muonium}. Computing the relevant matrix elements for the vector ortho-muonium state, we obtain the 
matrix elements
\begin{eqnarray}\label{ME1}
 \bra{\ammuv} Q_{1}\ket{\mmuv} &=& -3 f_M^2 M_M^2, \quad
 \bra{\ammuv} Q_{2}\ket{\mmuv} = -3 f_M^2 M_M^2, \nonumber \\ 
 \bra{\ammuv} Q_{3}\ket{\mmuv} &=& -\frac{3}{2} f_M^2 M_M^2, \quad
 \bra{\ammuv} Q_{4}\ket{\mmuv} = -\frac{3}{4} f_M^2 M_M^2, \\
 \bra{\ammuv} Q_{5}\ket{\mmuv} &=& -\frac{3}{4} f_M^2 M_M^2. \nonumber
\end{eqnarray}

Again, combining the contributions from the different operators, we obtain an expression for $x_V$ for the ortho-muonium state,
\begin{equation}\label{Delta_m_ortho}
x_V=-\frac{12 (m_{red}\alpha)^{3}}{\pi\Lambda^2 \Gamma}
\left[
C_1^{\Delta L= 2} + C_2^{\Delta L= 2} + \frac{1}{2} C_3^{\Delta L= 2} + 
\frac{1}{4}\left(C_4^{\Delta L= 2} + C_5^{\Delta L= 2}\right)
\right].
\end{equation}

The results in Equations (\ref{Delta_m_para}) and (\ref{Delta_m_ortho}) are universal and hold true for any new physics model that can be matched into a set of local $\Delta L = 2$ interactions. 

\subsection{The Lifetime Difference $y$}\label{LifeDiff}

The lifetime difference in the muonium system, defined in Equation (\ref{XandY}), is obtained from the absorptive part of Equation (\ref{OffDiagonal}) and
comes from the on-shell intermediate states common to both $\mmu$ and $\ammu$ \cite{Golowich:2006gq},
\beq\label{YPheno}
y = \frac{1}{\Gamma} \sum_{n} \rho_n \left \langle\ammu
\left|{\cal H}_{\rm eff} \right| n\right\rangle\left\langle n\left|{\cal H}_{\rm eff} \right| \mmu\right\rangle,
\eeq
where $\rho_n$ is a phase space function for a given intermediate state. There are many possible intermediate states composed of 
electrons, photons, and neutrinos. However, only the intermediate state containing neutrinos gives the largest contribution. 
This follows from the following argument. Noting that the contributions of multibody intermediate states are suppressed by the 
phase space factors $\rho_n$ for $n >2$, only two body intermediate states need to be considered. All possible SM two-body intermediate 
states that can contribute to $y$ are $e^+e^-$, $\gamma\gamma$, and $\nu\bar\nu$. 

The $e^+e^-$ intermediate state corresponds to a $\Delta L_{\mu}=1$ decay $\mmu \to e^+e^-$, which implies that 
${\cal H}_{\rm eff} = {\cal H}_{\rm eff}^{\Delta L_{\mu}=1}$ in Equation (\ref{YPheno}). According to Equation (\ref{L_eff1}), it appears that, 
quite generally, this contribution is suppressed by $\Lambda^4$, i.e., will be much smaller than $x$. The decays of muonia to the 
$\gamma\gamma$ intermediate states are generated by higher-dimensional operators and therefore are suppressed by even higher 
powers of $\Lambda$ or the QED coupling $\alpha$ than the contributions considered here. 

The only other possible contribution to $y$ comes from the on-shell $\nu\bar\nu$ intermediate state. This intermediate 
state can be reached by the standard model tree level decay $\mmu \to \overline{\nu_\mu} {\nu_e}$ and 
the $\Delta L_{\mu}=2$ decay $\ammu \to \overline{\nu_\mu} \nu_e$, i.e., it is common for both $\mmu$ and $\ammu$. 
This contribution is only suppressed by $\Lambda^2 M_W^2$ and represents the parametrically leading contribution to $y$ \cite{Conlin:2020veq}. 

Writing $y$ in terms of the absorptive part of the correlation function, 
\beqa\label{MEforY}
y &=& \frac{1}{2 M_M \Gamma} \mbox{Im} \left[ 
\langle\ammu \left| i\int d^4x \ \mbox{T} \left[
 {\cal H}_{\rm eff} (x) {\cal H}_{\rm eff} (0) \right] \right | \mmu\rangle 
\right]
\nonumber \\
&=& \frac{1}{M_M \Gamma} \mbox{Im} \left[ 
\langle\ammu \left| i\int d^4x \ \mbox{T} \left[
 {\cal H}_{\rm eff}^{\Delta L_\mu=2} (x) {\cal H}_{\rm eff}^{\Delta L_\mu=0} (0) \right] \right | \mmu\rangle
\right],
\eeqa
where the ${\cal H}_{\rm eff}^{\Delta L_\mu=0}=-{\cal L}_{\rm eff}^{\Delta L_\mu=0}$ is given by the ordinary standard model Lagrangian of
Equation (\ref{SMLagr}), and ${\cal H}_{\rm eff}^{\Delta L_\mu=2}$ only contributes through the operators $Q_6$ and $Q_7$ of Equation (\ref{Dim6_Op_nu}).

Since the decaying muon injects a large momentum into the two-neutrino intermediate state, the integral in Equation (\ref{MEforY})
is dominated by small distance contributions, compared to the scale set by $1/m_\mu$. We can compute the correlation 
function in Equation (\ref{MEforY}) by employing a short distance operator product expansion, systematically expanding it in 
powers of $1/m_\mu$ \cite{Conlin:2020veq}. 

\begin{figure}[H]
\center
\includegraphics[scale=1.0]{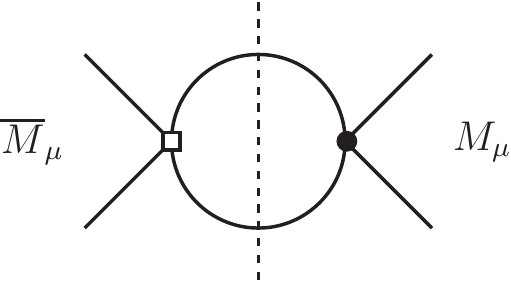}
\caption{A diagram, whose imaginary part, denoted by the dotted line, represents the lifetime difference $y$. 
A white square represents a ${\cal H}_{\rm eff}^{\Delta L_\mu=2}$, while a black dot is the 
SM contribution of Equation (\ref{SMLagr}).\label{Fig:DelGam}}
\end{figure}

Using Cutkoski rules to compute the discontinuity (imaginary part) of the transition amplitude (see Figure \ref{Fig:DelGam}), calculating 
the relevant phase space integrals, and taking the matrix elements for the spin-singlet and the spin-triplet states of the muonium 
we arrive at the lifetime differences for the two spin states \cite{Conlin:2020veq}.

{\bf Para-muonium}. The relevant matrix elements of the spin-singlet state can be read off the Equation (\ref{ME0}). 
Recalling the definitions in Equations (\ref{WFat0}) and (\ref{Definition_fM}), we obtain an expression for the lifetime difference $y_P$ 
for the para-muonium state,
\begin{equation}\label{Delta_Gamma_para}
y_P= \frac{G_F}{\sqrt{2} \Lambda^2} \frac{M_M^2}{\pi^2 \Gamma} (m_{red}\alpha)^{3} 
\left(C_6^{\Delta L= 2} - C_7^{\Delta L= 2}\right).
\end{equation}

One should note that if $C_6^{\Delta L= 2} = C_7^{\Delta L= 2}$ current conservation
assures that no lifetime difference is generated at this order in $1/\Lambda$ for the para-muonium.

{\bf Ortho-muonium}. Employing the matrix elements for the spin-triplet state computed in Equation (\ref{ME1}),
the lifetime difference for the vector muonia is 
\begin{equation}\label{Delta_Gamma_ortho}
y_V= -\frac{G_F}{\sqrt{2} \Lambda^2} \frac{M_M^2}{\pi^2 \Gamma} (m_{red}\alpha)^{3} 
\left( 5 C_6^{\Delta L= 2} + C_7^{\Delta L= 2}\right).
\end{equation}

We emphasize that Equations (\ref{Delta_Gamma_para}) and (\ref{Delta_Gamma_ortho}) represent 
parametrically leading contributions to muonium lifetime difference, as they are only suppressed by 
two powers of $\Lambda$. Nevertheless, additional suppression by $G_F$ makes the observation of the 
lifetime difference in the muonium system extremely challenging. 

\subsection{Experimental Studies of Muonium Oscillations}\label{Exp}

Both $x$ and $y$ are the observable parameters, so experiments studying $\mmu-\ammu$ oscillations
could probe them by producing the $\mmu$ state and looking for the decay products of the $\ammu$ state.
Such experiments must overcome several considerable challenges. 

First, muonium states must be 
produced inside the target and moved to the decay volume of the detector. The efficiency of such a process 
strongly depends on the target material. For example, the Muonium--Antimuonium Conversion Spectrometer 
(MACS) at PSI \cite{Willmann:1998gd} used SiO$_2$ powder target with efficiency of 4--5\%. A recently 
proposed Muonium-to-Antimuonium Conversion Experiment (MACE) \cite{MACE} at the China Spallation 
Neutron Source (CSNS) will use an aerogel target with laser-drilled channels. The efficiency of producing 
muonia on such target could reach 40\%.

Second, the strategy of looking for the ``wrong-sign" final state, while working well in the studies of $B^0$ or $D^0$ oscillations,
faces new challenges when applied to muonium oscillation searchers. The main reason is that the final state in
the muonium decay $\mmu \to e^+e^- \bar\nu_\mu \nu_e$ and that in the antimuonium decay $\ammu \to e^+e^- \bar\nu_e \nu_\mu$, 
differ only by the flavor combination of the neutrino states. Since the neutrinos are not detected, a method based on the
kinematics of the decay must be employed: the decay products of the $\ammu$ involve fast electron with the momentum of 
about 53 MeV and slow positron with the momentum of about 13.5 eV. This follows from the fact that the main decay channel of the 
state is the decay of a muon. 

Third, the experiments usually involve a setup where produced muonia propagate in the magnetic field $B_0$. 
This magnetic field suppresses oscillations by removing degeneracy between $\mmu$ and 
$\ammu$ \cite{Li:2019xvv,Hou:1995np}. It also has a different effect on different spin configurations of the 
muonium state and the Lorentz structure of the operators that generate mixing \cite{Cuypers:1996ia,Horikawa:1995ae}. 
Fortunately, these effects can be taken into account. MACS experiment corrected the oscillation probability by 
introducing a factor $S_B(B_0)$ \cite{Willmann:1998gd},
\beq\label{Exp_constr}
P(\mmu \rightarrow \ammu) \leq 8.3 \times 10^{-11}/S_B(B_0).
\eeq

The values of $S_B(B_0)$, presented in Table II of \cite{Willmann:1998gd},
are different for different values of magnetic field and chiral structure of the operators governing the 
$\mmbar$ oscillations. 

We can now use the derived expressions for $x$ and $y$ to place constraints on the BSM 
scale $\Lambda$ (or the Wilson coefficients $C_i$) from the experimental constraints on 
muonium--anti-muoium oscillation parameters. Since both spin-0 and spin-1 muonium states 
were produced in the experiment \cite{Willmann:1998gd}, we should average the oscillation 
probability over the number of polarization degrees of freedom, 
\beq
P(\mmu \rightarrow \ammu)_{\rm exp} = \sum_{i=P,V} \frac{1}{2S_i+1} P(\mmu^i \rightarrow \ammu^i),
\eeq
where $P(\mmu \rightarrow \ammu)_{\rm exp}$ is the experimental oscillation probability 
from Equation (\ref{Exp_constr}). We shall use the values of $S_B(B_0)$ for 
$B_0 = 2.8$ $\mu$T from the Table II of \cite{Willmann:1998gd}, as it will provide us the 
best experimental constraints on the BSM scale $\Lambda$.
We set the corresponding Wilson coefficient $C_i=1$.
We report those constraints in Table \ref{Table1}.
\begin{table}[H]
\caption{C\textls[-15]{onstraints on the energy scales probed by different $\Delta L=2$ operators of \mbox{Equations (\ref{Dim6_Op}) and (\ref{Dim6_Op_nu})}
in MACS experiment (from 
 \cite{Conlin:2020veq})}.}
\label{Table1}
\begin{tabularx}{\textwidth}{CCCC}
\toprule
\textbf{Operator} & \textbf{Interaction Type} & \boldmath{$ S_B(B_0)$} \textbf{(from \cite{Willmann:1998gd})} & \textbf{Scale} \boldmath{$\Lambda$}\textbf{, TeV} \\
\midrule
$Q_1$ & $(V-A)\times (V-A)$ & 0.75 & $5.4$ \\
$Q_2$ & $(V+A)\times (V+A)$ & 0.75 & $5.4$ \\
$Q_3$ & $(V-A)\times (V+A)$ & 0.95 & $5.4$ \\
$Q_4$ & $(S+P)\times (S+P)$ & 0.75 & $2.7$ \\
$Q_5$ & $(S-P)\times (S-P)$ & 0.75 & $2.7$ \\
$Q_6$ & $(V-A)\times (V-A)$ & 0.75 & $0.58\times 10^{-3}$ \\
$Q_7$ & $(V+A)\times (V-A)$ & 0.95 & $0.38\times 10^{-3}$ \\
\bottomrule
\end{tabularx}
 \end{table}
As one can see from Equations (\ref{Delta_m_para}), (\ref{Delta_m_ortho}), (\ref{Delta_Gamma_para}) and (\ref{Delta_Gamma_ortho}),
each observable depends on the combination of the operators. Assuming that only one operator at a time gives a dominant 
contribution, i.e., employing the single operator dominance hypothesis, it is possible to constrain the Wilson coefficients of each 
operator. While this ansatz is not necessarily realized in many
particular UV completions of the LFV EFTs, as cancellations among contributions of different operators are possible, it is, 
however, a useful tool in constraining parameters of ${\cal L}_{\rm eff}$.

We must emphasize that the constraints presented in Table \ref{Table1} use the data obtained more than twenty
years ago! New results from new experiments are therefore highly desired. The MACE experiment \cite{MACE} is 
expected to improve the sensitivity to $C_i^{\Delta L_\mu=2}/\Lambda^2$ by at least two orders of 
magnitude. A new experiment at J-PARC is also expected to improve the constraints on the muonium--antimuonium 
oscillation parameters \cite{Kawamura:2021lqk}.

\subsection{Constraints on Explicit Models of New Physics}\label{Models}

It might be instructive to consider explicit models of new physics which could be probed by $\mmu-\ammu$ oscillations. 
This approach lacks the universality of the EFT approach described above. However, what it lacks in the universality 
it compensates for in the applicability: new degrees of freedom introduced in explicit models can contribute to other 
processes, some of which might not even include FCNCs or muons. Even restricting our attention to the sector of those
frameworks containing $\Delta L_\mu=1$ and/or $\Delta L_\mu=2$ interactions reveals such a multitude of models that 
reviewing each and every one of them here would be impractical. We will take another approach. 

We will review two specific models, one with heavy NP particles (a doubly-charged Higgs model), and one with light 
NP states (a model with flavor-violating axion-like particle (ALPs)) as examples. Then we discuss classes of NP interactions
that can be probed by muonium oscillations, and refer the readers to further examples.

One way to classify the models of NP that contribute to $\mmu-\ammu$ mixing is by the masses of NP particles $m$. 
The models with heavy ($m \gg m_\mmu$) new physics degrees of freedom can be matched to the effective
Lagrangian of Equation (\ref{DL2}). With that, experimental constraints on the parameters of this Lagrangian discussed in 
Section \ref{Exp} and Table \ref{Table1} lead to constraints on model parameters. The constraints on models with 
light ($m \leq m_\mmu$) new physics degrees of freedom can be implemented within the framework of either concrete 
or simplified models.

{\bf Heavy new physics.} Let us consider a model which contains a doubly-charged Higgs boson \cite{Swartz:1989qz,Chang:1989uk,Han:2021nod}. Such states often appear in the context of
left-right models \cite{Kiers:2005gh,Kiers:2005vx}, where an additional Higgs triplet is introduced to introduce 
neutrino masses
\beq
\Delta =
 \left( {\begin{array}{cc}
 \Delta^+/\sqrt{2} & \Delta^{++} \\
 \Delta^0 & -\Delta^+/\sqrt{2} \\
 \end{array} } \right)
\eeq

A coupling of the doubly charged Higgs field $\Delta^{++}$ to the lepton fields 
can be written as
\beq
{\cal L}_R = -\frac{1}{2} g_{\ell_i \ell_j} \overline \ell_i^c \Delta \ell_j + H.c.,
\eeq
where $\ell^c=C\overline{\ell}^T$ is the charge-conjugated lepton state. Integrating out the $\Delta^{--}$ field,
this Lagrangian leads to the following effective Hamiltonian \cite{Swartz:1989qz,Kiers:2005vx}
\beq\label{DoublyChargedHiggs}
{\cal H}_\Delta = -\frac{g_{ee} g_{\mu\mu}^*}{8 M_\Delta^2} \left(\overline\mu_L \gamma_\alpha e_L \right) 
\left(\overline\mu_L \gamma^\alpha e_L \right) + H.c.,
\eeq
below the scales associated with the doubly-charged Higgs field's mass $M_\Delta$. Examining 
Equation (\ref{DoublyChargedHiggs}) we see that this Hamiltonian matches onto our operator $Q_2$ 
(see Equation (\ref{Dim6_Op})) with the scale $\Lambda = M_\Delta$ and the corresponding Wilson coefficient
$C_2^{\Delta L= 2} = g_{ee} g_{\mu\mu}^*/8$. 

The constraints on the masses and coupling constants depend on a particular model and other 
assumptions, such as whether the hierarchy of the neutrino masses is direct or inverse. It is however 
claimed that future experiments, such as MACE, could provide constraints on the 
doubly-charged Higgs state mass of $m_\Delta < 3$ TeV \cite{Han:2021nod}. 

It is interesting to point out that such a doubly-charged scalar state would also contribute to the anomalous magnetic moment
of the muon, both at one-loop \cite{Fukuyama:2009xk} and at two-loops via the Barr-Zee type of mechanism \cite{Chen:2021jok}. 
It was shown that $\Delta^{++}$ with the mass of a few hundred GeV could explain the discrepancy between the 
theoretical prediction and experimental measurement of $(g-2)$ of the muon \cite{Muong-2:2021ojo}, particularly due to the 
enhancement from the two units of electric charge of the $\Delta^{++}$. 

{\bf Light new physics.} Let us consider a model with light axion-like particles that couple derivatively to the 
lepton current \cite{Endo:2020mev,Calibbi:2020jvd}. For the flavor off-diagonal interactions of the ALP $a$, 
the Lagrangian would contain a term
\begin{equation}\label{ALPinteraction}
{\cal L}_{\rm ALP} = \sum_{i\neq j} \frac{\partial_\mu a}{2 f_a} \ \bar\ell_i \gamma^\mu 
\left(C_{\ell_i\ell_j}^V + C_{\ell_i\ell_j}^A \gamma_5\right) \ell_j,
\end{equation}
where $C_{\ell_i\ell_j}^{V,A}$ are Hermitian matrices of coupling constants, and $f_a$ is a decay constant related to
the scale of the symmetry breaking associated with the ALP. 

Since the mass of the ALP, $m_a$, is a free parameter, it is possible that, while $a$ is light, to have $m_a > m_\mmu$. In this 
case the ALP can be integrated out and the constraints from the Table \ref{Table1} imply that \cite{Calibbi:2020jvd}
\begin{equation}\label{ALPheavy}
\frac{1}{3.8 \ \mbox{TeV}} > \frac{1}{2 f_a} \left| \left(C_{\mu e}^A\right)^2 - 
\left(C_{\mu e}^V\right)^2\right|^{1/2} 
\left(\frac{m_\mu}{m_e} \right). 
\end{equation}

In the opposite case $m_a < m_\mmu$ the constraint can be obtained by taking a limit $m_\mu/m_a \to 1$ \cite{Calibbi:2020jvd}.
Note that in both cases one can only constrain the combination $C_{\mu e}^{V,A}/f_a$. These results have direct implications 
for ALP contributions to muon $(g-2)$ \cite{Endo:2020mev} ({c.f.,} \cite{Bauer:2019gfk}), the cross-section of 
$e^+ e^- \to ee\mu\mu$ \cite{Calibbi:2020jvd}, and other quantities. 

A useful classification of lepton-flavor-violating interactions probed by the muonium oscillations was given 
in \cite{Fukuyama:2021iyw}. Such interactions can be classified by the way they break the lepton flavor. 
\begin{enumerate}
\item
$\Delta L_e = \Delta L_\mu = 0$. In such models, the interactions do not violate lepton flavor quantum numbers. 
The muonium oscillations are induced at one-loop order by the mass terms of the fields that transform as singlets 
under the SM gauge group and have $L_{e,\mu} = \pm 2$. Examples of such models include constructions with Majorana 
neutrinos \cite{ClarkLove:2004,Fukuyama:2021iyw}.
\item
$\Delta L_e = \pm 2, \Delta L_\mu = 0$ and $\Delta L_e = 0, \Delta L_\mu = \pm 2$. In such models, the lepton
flavor quantum numbers are separately broken. The mediators of such interactions, sometimes called dilepton 
bosons, have electric charge $2e$ and lepton number equal to two. The mediators could be scalar or vector bosons. 
Examples of such models include the doubly-charged Higgs model considered above \cite{Cuypers:1996ia,Horikawa:1995ae,Swartz:1989qz,Chang:1989uk,Han:2021nod,Fukuyama:2009xk,Chen:2021jok} 
and many other constructions \cite{Fukuyama:2021iyw}. Muonium oscillations can be generated at tree level in 
such models.
\item
$\Delta L_e = -\Delta L_\mu = \pm 1$. In such models, interaction terms violate both lepton flavor numbers. The mediators of 
such interactions are electrically neutral and could be both scalar and vector bosons. Examples of such models include models
with flavor-violating Higgs boson \cite{Harnik:2012pb}, and many others \cite{Fukuyama:2021iyw}. Muonium oscillations can also be 
generated at the tree level in such models. These models can also be probed in muon conversion experiments unless the 
mediator is introduced such that it only interacts with the leptons. 
\item
$\Delta L_e = \pm 1, \Delta L_\mu = 0$ and $\Delta L_e = 0, \Delta L_\mu = \pm 1$. In such models, effective operators 
mediating muonium oscillations are generated at one-loop order. These models can also be probed in other 
muon transitions, such as $\mu \to 3e$. 
\end{enumerate}

As was pointed out in \cite{Fukuyama:2021iyw} if a $Z_n$ discrete symmetry is imposed to suppress the 
$\Delta L_e -\Delta L_\mu = \pm 1$ interactions, while allowing for $\Delta L_e -\Delta L_\mu = \pm 2$ terms, 
the oscillation rates are not well-probed by other experiments and can be as large as allowed by the current
experimental bound. Clearly, $\mmu-\ammu$ oscillations probe a wide variety of NP models and serve as effective 
tools that are complementary to other searches. 

\section{Muonium Decays}

Processes that change lepton flavor by two units can, in principle, also be studied in the decays of muons and muonia. 
The transitions into neutrino states, governed by the operators in Equation (\ref{Dim6_Op_nu}), fit the bill. The decay rate 
for such transition is
\beq\label{DecayBSM}
\Gamma\left(\mmuv \to \bar \nu_e \nu_\mu \right) = 
\frac{f_M^2 M_M^3}{9\pi \Lambda^4} 
\left|C_6^{\Delta L_\mu=2} + C_7^{\Delta L_\mu=2} \right|^2.
\eeq

Note that in the limit of massless neutrinos only the
spin-one ortho-muonium state would have non-zero width. While the invisible decays of the spin-zero para-muonium
would not be strictly forbidden in this limit, as they would be dominated by the four-neutrino final state \cite{Bhattacharya:2018msv}, 
the decay rate of the para-muonium would be very small. 

The decay rate of Equation (\ref{DecayBSM}) will add incoherently to the SM decay rate 
\beq\label{DecaySM}
\Gamma\left(\mmuv \to \bar \nu_\mu \nu_e \right) = 
\frac{G_F^2 f_M^2 M_M^3}{12\pi}.
\eeq

Even in the SM, this is a very rare process. Numerically, the branching fraction can be computed to be
\beq
{\cal B} \left(\mmu \to \bar \nu_\mu \nu_e \right) = 6.6 \times 10^{-12}.
\eeq

Both decays contribute to the ``invisible" width of the muonium,
\beq\label{InvDec}
\Gamma\left(\mmuv \to \mbox{invisible} \right) = 
\Gamma\left(\mmuv \to \bar \nu_\mu \nu_e \right) + 
\Gamma\left(\mmuv \to \bar \nu_e \nu_\mu \right) + ...,
\eeq
where ellipses denote decays into multineutrino states, which are further suppressed by the phase space factors. 

Experimental studies of the invisible width of muonia are very challenging. One possibility to 
constrain the invisible width indirectly \cite{Gninenko:2012nt} involves comparing the measurements of the
positively-charged muon lifetime inside the material, where it forms the muonia and that of the free muon. 
Since the lifetime difference between the muonium in the target and free muon in vacuum was estimated to be negligible
 \cite{Czarnecki:1999yj}, the possible difference can be ascribed to the invisible width of the muonium. Using the
lifetime measurements performed by the MuLan Collaboration \cite{MuLan:2007qkz}, the limit 
\beq
{\cal B} \left(\mmu \to \mbox{invisible} \right) < 5.7 \times 10^{-6} 
\eeq
can be established at 90\% C.L. \cite{Gninenko:2012nt} with an assumption that the fraction of triplet muonium state in the 
MuLan's quartz target is 3/4. This bound does not yet provide a meaningful probe of either an SM process of Equation (\ref{DecaySM}) 
or the $\Delta L_\mu=2$ process of Equation (\ref{DecayBSM}). 

A more sensitive probe of the $\Delta L_\mu=2$ process does not require a muonium bound state and can be obtained by 
comparing the lifetime of the muon in a vacuum to theoretical prediction in the framework of the standard model. The results will
be reported elsewhere.

\section{Conclusions}

Muonium is the simplest QED bound state. Yet, it holds the potential to probe new physics in 
ways that are unique and complementary to other NP searches with muons. In this review, we discussed how 
lepton-flavor violating $\Delta L_\mu=2$ transitions could be probed in the muonium system, both in 
muonium decays and in muonium--antimuonium oscillations. We discussed the phenomenology of
such oscillations, arguing that the presence of $\Delta L_\mu=2$ interactions would lead to both
mass and lifetime splittings in the muonium--antimuonium system. We showed how to compute 
those oscillation parameters in an effective field theory approach, showing that both
parameters scale as $1/\Lambda^2$ with respect to the new physics scale $\Lambda$.

We discussed the current status and future perspectives of experiments that can constrain 
muonium--antimuonium oscillation parameters, and how such measurements can be used to
place bounds on the parameters of the effective Lagrangians governing the $\Delta L_\mu=2$ 
transitions. These bounds can be further translated into the constraints on the masses and couplings 
of new particles in specific models of new physics, and, in case of the observation of the
oscillation phenomena would help us to identify the proper extension of the standard model. 

\vspace{6pt} 





\funding{This research was supported in part by the U.S. Department of Energy grant No. DE-SC0007983 and by the Excellence Cluster ORIGINS, which is 
funded by the Deutsche Forschungsgemeinschaft (DFG, German Research Foundation) under Germany's Excellence Strategy ---EXC-2094---390783311. 
AAP was also supported by the Visiting Scholars Award Program of the Universities Research Association.
Fermilab is managed and operated by Fermi Research Alliance, LLC under Contract 
No. DE-AC02-07CH11359 with the U.S. Department of Energy.}

\institutionalreview{Not applicable.
}

\informedconsent{Not applicable. 

}

\dataavailability{Not applicable.
} 

\acknowledgments{A.A.P. thanks Pittsburgh Particle Physics Astrophysics and Cosmology Center (Pitt PACC), where this paper
was partially completed, for its kind hospitality and stimulating research environment.}

\conflictsofinterest{The authors declare no conflict of interest.} 



\begin{adjustwidth}{-\extralength}{0cm}
\printendnotes[custom]
\end{adjustwidth}
\reftitle{References}




\begin{adjustwidth}{-\extralength}{0cm}

\end{adjustwidth}

%


\end{document}